# Electron–photon entanglement in real space from quantum electrodynamics

**Ivan P. Christov**

Physics Department, Sofia University, 1164 Sofia, Bulgaria

**Abstract:** We study the entanglement between an electron and the photon it scatters, resolved in real space and time. Starting from nonrelativistic QED in Fermi's oscillator representation, we write Fock's coupled equations for the components with fixed photon number in the configuration space of the photons, in the length gauge, where the photon is created around the atom with a nonlocal kernel. Truncated at one photon, the equations are solved numerically, without the rotating-wave approximation, for a one-dimensional model atom in a waveguide with a single transverse mode and an enhanced coupling, driven by a short off-resonant laser pulse, with the electron and the photon on real-space grids. The global electron–photon entanglement is largely transient. It peaks during the pulse and decreases when the polarized electron cloud returns to the core, leaving a state close to a product of the electron state and a superposition of the field vacuum with one scattered photon. Local entanglement, obtained from walkers sampled over both photon-number components, shows opposite patterns for the two particles. The electron entropy is smallest at the nucleus and maps the local strength of the dipole coupling, while the photon entropy is largest near the atom, where photons are created and absorbed, and smaller in the outgoing packets.

## 1. Introduction

The interaction of light with matter is usually described on one of two levels. In the semiclassical picture the electrons are quantum particles while the field obeys Maxwell's equations with the mean electron current as its source. In quantum electrodynamics (QED) the field is quantized as well [1, 2, 3]. With self-consistent back-reaction the semiclassical picture accounts for absorption, stimulated emission and coherent scattering, but it fails for processes in which the number of photons matters. An atom prepared in a pure excited state never decays because its mean dipole vanishes, a single photon does not give coincident counts at two separated detectors [4], and resonance fluorescence shows no antibunching [5]. In all these processes the state of the electrons and the state of the field are entangled, and it is this electron–photon entanglement that we want to follow in real space and time.

The standard route to QED expands the transverse field in modes, each a harmonic oscillator. Numerical methods for atoms and molecules in cavities follow this route, either exactly in a truncated photon-number space or approximately, as in QED Hartree–Fock and coupled-cluster theory [6] and in diffusion Monte Carlo with the mode coordinate as an additional walker coordinate [7]. In these methods the position of a photon is not a variable, and the cost grows with the number of modes rather than with the number of photons. Entanglement in real space requires a description in which the photon has a coordinate of its own.

Such a description goes back to Landau and Peierls [8], who wrote QED as a Schrödinger equation for a sequence of wave functions with $0,1,2,\ldots$ photons in configuration space. Fock showed that these sequences are equivalent to second quantization [9] and formulated QED as coupled equations for the components with fixed photon number [10]. When truncated at one photon, they give the Breit interaction and the natural line width. The real-space photon amplitude with a local probability density is the number, or Landau–Peierls, amplitude [11, 12, 13, 14]. It differs from the energy amplitude, the positive-frequency Riemann–Silberstein field normalized to one quantum, by the nonlocal factor $(\hbar c\,|\nabla|)^{1/2}$, which reflects the absence of localized photon states and enters the coupling of the photon to the charges.

Real-space photons coupled to emitters are standard in waveguide QED [15, 16], mostly for two- or three-level emitters within the rotating-wave approximation (RWA). Atom–photon entanglement in real space has been obtained from the Wigner–Weisskopf amplitudes of two-level atoms with recoil [17, 18], and numerical solutions exist for two-level atoms coupled to many cavity modes in the space of a single excitation [19, 20]. The single-mode form of Fock's equations, with the driving laser as the quantized mode and in the velocity gauge, has been solved for strong-field ionization [21], where the entanglement arises from the spread in the number of photons absorbed from the drive, which contains up to $N_0 \approx 10^6$ photons, and spontaneous emission is neglected as a correction of order $1/N_0$. In the quantum optics of strong-field processes the coherent drive is split into a classical field and its quantum fluctuations, and all modes start from the vacuum. When the field couples only to the mean dipole, it ends in a product of coherent states, and nonclassical and entangled states of light are obtained by conditioning on high-harmonic generation or above-threshold ionization [22]. These treatments use field modes and do not resolve the photon in space. To our knowledge, the photon-number equations have not been solved for an electron on a real-space grid, including its continuum, with the photon in position space and without the RWA.

In this work we use waves and walkers in a quantum Monte Carlo approach [23, 24, 25], but the conditional waves are slices of the exact numerical solution of the Schrödinger equation at the positions of the walkers. In the configuration-space form of QED a walker carries an electron coordinate and, in the one-photon component, also a photon coordinate. Grouping the walkers by their positions, we define the entanglement locally, since the reduced density matrix of a spatial region is built from the conditional waves of the walkers located in it [26].

We first write Fock's photon-number hierarchy in the configuration space of the photons, in the length gauge, starting from Fermi's oscillator representation of the field; the photon is created around the atom with the kernel $|\nabla|^{1/2}\delta^{\perp}$, and the truncation at one photon fixes how the polarization energy enters each component. Second, we solve the truncated equations without the RWA for an electron on a real-space grid coupled to a continuum of guide modes. Third, we apply the strip construction of local entanglement to both particles, with the walkers distributed in the configuration space of the electron–photon system. Section 2 contains the derivation and the waveguide model, and Section 3 the numerical method and the results for an atom driven by a short off-resonant laser pulse.

## 2. Nonrelativistic QED in the configuration space of photons

### 2.1. Electrons and the transverse field

We consider $N$ electrons of charge $-e$ and mass $m$ at the positions $\mathbf{R} \equiv (\mathbf{r}_1, \dots, \mathbf{r}_N)$, bound to a nucleus of charge $Ze$ held fixed at $\mathbf{X}$. We split the electric field into its longitudinal and transverse parts, $\mathbf{E} = \mathbf{E}_\parallel + \mathbf{E}_\perp$, with $\nabla \times \mathbf{E}_\parallel = 0$ and $\nabla \cdot \mathbf{E}_\perp = 0$. Gauss's law, $\nabla \cdot \mathbf{E}_\parallel = 4\pi\rho$, fixes $\mathbf{E}_\parallel$ at every instant by the positions of the charges, and its energy $(1/8\pi)\int \mathbf{E}_\parallel^2 \, d^3x$ is the Coulomb energy. After the infinite self-energies of the point charges are dropped, this energy becomes the potential

$$V_C(\mathbf{R}) = -\sum_{i=1}^{N} \frac{Ze^2}{|\mathbf{r}_i - \mathbf{X}|} + \sum_{i<j} \frac{e^2}{|\mathbf{r}_i - \mathbf{r}_j|} \tag{1}$$

In the Coulomb gauge, $\nabla \cdot \mathbf{A} = 0$, the vector potential $\mathbf{A}$ is purely transverse, $\mathbf{E}_\perp = -c^{-1}\,\partial_t \mathbf{A}$ and $\mathbf{B} = \nabla \times \mathbf{A}$. The Hamiltonian of the electrons and the field is [27]

$$H = \sum_{i=1}^{N} \frac{1}{2m}\left(\mathbf{p}_i + \frac{e}{c}\mathbf{A}(\mathbf{r}_i)\right)^2 + V_C(\mathbf{R}) + \frac{1}{8\pi}\int (\mathbf{E}_\perp^2 + \mathbf{B}^2) d^3x \tag{2}$$

where $\mathbf{p}_i$ is the canonical momentum of electron $i$. The minimal coupling in the first term gives the Lorentz force on the electrons and, through the same action, the transverse current as the source of the field. The photons are the quanta of the transverse dynamical field.

### 2.2. Dipole approximation and the length gauge

We place the field in a periodic box of volume $V$ and expand it in a complete set of real, orthonormal, transverse mode functions $\mathbf{u}_\nu(\mathbf{x})$, with $\nabla \cdot \mathbf{u}_\nu = 0$, $-\nabla^2 \mathbf{u}_\nu = k_\nu^2 \mathbf{u}_\nu$ and $\int \mathbf{u}_\nu \cdot \mathbf{u}_\mu \, d^3x = \delta_{\nu\mu}$. In free space these are standing waves $\cos(\mathbf{k}\cdot\mathbf{x})$ and $\sin(\mathbf{k}\cdot\mathbf{x})$ times two polarization vectors perpendicular to $\mathbf{k}$. Their completeness relation defines the transverse delta function, with the Cartesian components labelled by the Greek indices $\alpha, \beta$,

$$\sum_\nu u_\nu^\alpha(\mathbf{x})\, u_\nu^\beta(\mathbf{x}') = \delta_{\alpha\beta}^\perp(\mathbf{x} - \mathbf{x}') \tag{3}$$

With the real amplitudes $Q_\nu(t)$ we write

$$\mathbf{A}(\mathbf{x}) = \sqrt{4\pi}\, c \sum_\nu Q_\nu \, \mathbf{u}_\nu(\mathbf{x}), \qquad \mathbf{E}_\perp(\mathbf{x}) = -\sqrt{4\pi} \sum_\nu \dot{Q}_\nu \, \mathbf{u}_\nu(\mathbf{x}) \tag{4}$$

Inserting these into the field energy, and using $\int (\nabla \times \mathbf{u}_\nu) \cdot (\nabla \times \mathbf{u}_\mu)\, d^3x = k_\nu^2 \delta_{\nu\mu}$, we obtain a sum of oscillators of unit mass with the frequencies $\omega_\nu = ck_\nu$,

$$\frac{1}{8\pi}\int (\mathbf{E}_\perp^2 + \mathbf{B}^2) d^3x = \sum_\nu \frac{1}{2}(P_\nu^2 + \omega_\nu^2 Q_\nu^2) \tag{5}$$

where $P_\nu = \dot{Q}_\nu$ is the momentum conjugate to $Q_\nu$. This is Fermi's form of the radiation field [3]. We quantize it in the Schrödinger representation, with one wave function $\Psi(\mathbf{R}, \{Q_\nu\}, t)$ of all electron coordinates and all field amplitudes, $\mathbf{p}_i = -i\hbar\nabla_i$ and $P_\nu = -i\hbar\, \partial/\partial Q_\nu$. The free field Hamiltonian is

$$H_F = \sum_\nu \frac{1}{2}\left(-\hbar^2 \frac{\partial^2}{\partial Q_\nu^2} + \omega_\nu^2 Q_\nu^2\right) - E_{zp}, \qquad E_{zp} = \sum_\nu \frac{\hbar\omega_\nu}{2} \tag{6}$$

The constant $E_{zp}$, the zero-point energy of the field oscillators, is subtracted, and all energies are counted from the field vacuum, the state in which every oscillator is in its ground state. The eigenfunctions of each oscillator are the normalized Hermite functions $\eta_n(Q_\nu)$ with the energies $n\hbar\omega_\nu$ above the vacuum, and $n$ is the number of photons in mode $\nu$.

We write $\mathbf{y}_i = \mathbf{r}_i - \mathbf{X}$ for the electron coordinates relative to the nucleus, and $\mathbf{d} = -e\sum_i \mathbf{y}_i$ for the dipole of the neutral atom ($Z = N$). The electrons are bound within a few Bohr radii, while the wavelengths that couple to them are much longer. In the dipole approximation we replace $\mathbf{A}(\mathbf{r}_i)$ by $\mathbf{A}(\mathbf{X})$. For an atom of size $\ell$ the neglected electric-quadrupole and magnetic-dipole couplings are of relative order $(k\ell)^2$ for the coupled wave numbers $k$. The Hamiltonian becomes

$$H_v = \sum_i \frac{1}{2m}\left(\mathbf{p}_i + \frac{e}{c}\mathbf{A}(\mathbf{X})\right)^2 + V_C(\mathbf{R}) + H_F \tag{7}$$

Since $\mathbf{A}(\mathbf{X})$ is a linear function of the field amplitudes, the transformation to the length gauge [28] is in Fermi's representation simply the multiplication of the wave function by a phase that depends on the electron and field coordinates:

$$\Psi = \exp\left(\frac{i\sqrt{4\pi}}{\hbar}\sum_\nu (\mathbf{d}\cdot\mathbf{u}_\nu(\mathbf{X}))\, Q_\nu\right)\Psi' \tag{8}$$

The phase shifts the electron momenta by $-(e/c)\mathbf{A}(\mathbf{X})$, which removes the vector potential from the kinetic energy, and it shifts each field momentum by $P_\nu \to P_\nu + \sqrt{4\pi}\,\mathbf{d}\cdot\mathbf{u}_\nu(\mathbf{X})$. Expanding the field energy with the shifted momenta, we arrive at the length-gauge Hamiltonian

$$H' = H_{el} + H_F - \mathbf{d}\cdot\mathbf{D}_\perp(\mathbf{X}) + V_P(\mathbf{R}), \qquad H_{el} = \sum_i \frac{\mathbf{p}_i^2}{2m} + V_C(\mathbf{R}) \tag{9}$$

with

$$\mathbf{D}_\perp(\mathbf{x}) = -\sqrt{4\pi}\sum_\nu P_\nu\ \mathbf{u}_\nu(\mathbf{x}), \qquad V_P(\mathbf{R}) = 2\pi\sum_\nu (\mathbf{d}\cdot\mathbf{u}_\nu(\mathbf{X}))^2 = 2\pi\, d_\alpha\, \delta^\perp_{\alpha\beta}(0)\, d_\beta \tag{10}$$

The field that couples to the dipole is the transverse displacement field $\mathbf{D}_\perp = \mathbf{E}_\perp + 4\pi\mathbf{P}_\perp$, with the polarization $\mathbf{P} = \mathbf{d}\,\delta(\mathbf{x}-\mathbf{X})$; outside the atom it coincides with the electric field [27, 29]. The term quadratic in $\mathbf{A}$ has disappeared, and in its place appears the transverse polarization energy $V_P = 2\pi\int |\mathbf{P}_\perp|^2\, d^3x$, often called the dipole self-energy [30]. It is not added to the Hamiltonian. Writing the radiation energy of Eq. (2) with $\mathbf{E}_\perp = \mathbf{D}_\perp - 4\pi\mathbf{P}_\perp$ gives

$$\frac{1}{8\pi}\int (\mathbf{E}_\perp^2 + \mathbf{B}^2) d^3x = \frac{1}{8\pi}\int (\mathbf{D}_\perp^2 + \mathbf{B}^2) d^3x - \mathbf{d}\cdot\mathbf{D}_\perp(\mathbf{X}) + V_P \tag{11}$$

and $V_P$ is the part of the transverse field energy carried by the polarization of the atom. Together with the coupling it completes a square in each field oscillator, $H' = H_{el} + \sum_\nu \frac{1}{2}\Big[\big(P_\nu + \sqrt{4\pi}\,\mathbf{d}\,\cdot$ $\mathbf{u}_\nu(\mathbf{X})\big)^2 + \omega_\nu^2 Q_\nu^2\Big] - E_{zp}$, which makes $H'$ bounded from below; without it the length-gauge Hamiltonian has no ground state [30].

### 2.3. Components with fixed photon number

The coupling in Eq. (9) is linear in the field momenta. In the Schrödinger representation of the field, the momentum acts on the Hermite functions as

$$P_\nu\, \eta_n(Q_\nu) = -i\sqrt{\frac{\hbar\omega_\nu}{2}}\left(\sqrt{n}\,\eta_{n-1}(Q_\nu) - \sqrt{n+1}\,\eta_{n+1}(Q_\nu)\right) \tag{12}$$

which follows from the recursion relations of the Hermite polynomials. Each action of the coupling therefore changes the number of photons by one. We expand $\Psi'(\mathbf{R},\{Q_\nu\},t)$ in products of oscillator eigenfunctions and collect the terms with the total photon number $n$. Following Fock [9] we write the coefficients of these terms as functions $\psi_n(\mathbf{R};\nu_1,\dots,\nu_n,t)$ that are symmetric in the mode labels and antisymmetric in the electron coordinates; a product state with $n_\nu$ photons in mode $\nu$ enters with the coefficient $(n!/\prod_\nu n_\nu\,!)^{1/2}\,\psi_n$. With this convention the normalization takes the form

$$\sum_{n=0}^{\infty}\ \sum_{\nu_1\dots\nu_n} \int |\psi_n(\mathbf{R};\nu_1,\dots,\nu_n,t)|^2\, d\mathbf{R} = 1 \tag{13}$$

and $|\psi_n|^2$ is the probability density of finding the electrons at $\mathbf{R}$ together with $n$ photons in the modes $\nu_1,\dots,\nu_n$. We introduce the field of one photon in mode $\nu$ at the nucleus,

$$\mathbf{g}_\nu = \sqrt{2\pi\hbar\omega_\nu}\ \mathbf{u}_\nu(\mathbf{X}) \tag{14}$$

which gives $V_P = \sum_\nu (\,\mathbf{d}\cdot\mathbf{g}_\nu)^2/\hbar\omega_\nu$. Inserting the expansion into $i\hbar\,\partial_t\Psi' = H'\Psi'$, using Eq. (12) and projecting onto the oscillator states, we obtain the hierarchy

$$\begin{aligned} i\hbar\,\partial_t\psi_n = {} & \left[H_{el} + V_P + \sum_{l=1}^{n}\hbar\,\omega_{\nu_l}\right]\psi_n - i\sqrt{n+1}\sum_\nu (\,\mathbf{d}\cdot\mathbf{g}_\nu)\,\psi_{n+1}(\nu,\nu_1,\dots,\nu_n) \\ & + \frac{i}{\sqrt{n}}\sum_{l=1}^{n}(\,\mathbf{d}\cdot\mathbf{g}_{\nu_l})\,\psi_{n-1}(\nu_1,\dots,\hat{\nu}_l,\dots,\nu_n) \end{aligned} \tag{15}$$

where $\hat{\nu}_l$ means that the label $\nu_l$ is left out. The second term on the right is the absorption of one photon from the component with $n+1$ photons, the third term the emission of one photon from the component with $n-1$ photons. Eq. (15) has the structure of Fock's coupled equations, here in the length gauge. Creation and annihilation operators are not needed for it, as they are only a shorthand for Eq. (12). The oscillator amplitudes $Q_\nu$ appear only in the derivation and are absent from the result.

### 2.4. Photons in real space

For each photon we transform the mode label into a position $\mathbf{x}$ and a Cartesian index $\alpha$,

$$\phi_n^{\alpha_1\dots\alpha_n}(\mathbf{R};\mathbf{x}_1,\dots,\mathbf{x}_n,t) = \sum_{\nu_1\dots\nu_n} u_{\nu_1}^{\alpha_1}(\mathbf{x}_1)\cdots u_{\nu_n}^{\alpha_n}(\mathbf{x}_n)\ \psi_n(\mathbf{R};\nu_1,\dots,\nu_n,t) \tag{16}$$

Since the $\mathbf{u}_\nu$ form an orthonormal basis of the transverse fields, Eq. (16) is a unitary map onto transverse functions of each photon coordinate, and the normalization (13) becomes the integral of $\sum_n |\phi_n|^2$ over all electron and photon coordinates. The components $\phi_n$ are the number amplitudes of the photons, and $|\phi_n|^2$ is the joint probability density of the electrons and of $n$ photons in configuration space. The photon energy $\sum_l \hbar\,\omega_{\nu_l}$ becomes the operator $\sum_l \hbar\, c\,|\nabla_l|$, where $|\nabla| = (-\nabla^2)^{1/2}$ multiplies each plane wave $e^{i\mathbf{k}\cdot\mathbf{x}}$ by $|\mathbf{k}|$. The coupling constants $\mathbf{g}_\nu$ combine into the kernel

$$G_{\alpha\beta}(\mathbf{x}-\mathbf{X}) = \sum_\nu u_\nu^\alpha(\mathbf{x})\, g_\nu^\beta = \sqrt{2\pi\hbar c}\,\left[|\nabla|^{1/2}\delta^\perp_{\alpha\beta}\right](\mathbf{x}-\mathbf{X}) \tag{17}$$

which is real, even and transverse. $G(\mathbf{x}-\mathbf{X})$ is the field at the nucleus that belongs to one photon at $\mathbf{x}$; read in the other direction, it is the shape with which a photon is created around the nucleus. Away from the origin $\delta^\perp$ has the form of a static dipole field, $\propto |\mathbf{x}|^{-3}$, and the factor $|\nabla|^{1/2}$ steepens it to a tail $|\mathbf{x}|^{-7/2}$. The emitted photon is born with a power-law tail around the atom, not at a point; this is where the photon localization problem enters the dynamics. Transforming Eq. (15) photon by photon we obtain the hierarchy in real space:

$$\begin{aligned} i\hbar\,\partial_t\phi_n = {} & \left[H_{el} + V_P + \sum_{l=1}^{n} \hbar\, c\,|\nabla_l|\right]\phi_n - i\sqrt{n+1}\;\mathbf{d}\cdot\int G(\mathbf{x}-\mathbf{X})\cdot\phi_{n+1}(\mathbf{R};\mathbf{x},\mathbf{x}_1,\dots,\mathbf{x}_n)\,d^3x \\ & + \frac{i}{\sqrt{n}}\sum_{l=1}^{n} G\,(\mathbf{x}_l-\mathbf{X})\cdot\mathbf{d}\;\phi_{n-1}(\mathbf{R};\mathbf{x}_1,\dots,\hat{\mathbf{x}}_l,\dots,\mathbf{x}_n) \end{aligned} \tag{18}$$

In the absorption term the vector index of the absorbed photon is contracted with $G$, in the emission term the new photon carries the index of $G$. Up to the dipole approximation, Eq. (18) is the Schrödinger equation of nonrelativistic QED in the configuration space of $N$ electrons and a variable number of transverse photons. Several of its features are needed below.

(i) *The vacuum.* The bare field vacuum is the component $\phi_0$ without photons; it carries no photon coordinate and is not represented by photon walkers. Vacuum fluctuations of the free field do not appear because energies are counted from the vacuum. Spontaneous emission and the radiative level shifts are nevertheless contained in Eq. (18), in the form in which they are attributed to the reaction of the atom to its own field [31].

(ii) *The source.* The emission term is driven by the amplitude $\mathbf{d}(\mathbf{R})\,\phi_{n-1}(\mathbf{R})$, by the transition dipole of the electrons, and not by the mean dipole $\langle\mathbf{d}\rangle$, which vanishes for an eigenstate of the atom. This is the difference from the semiclassical equations with back-reaction, in which an atom in an excited eigenstate does not radiate.

(iii) *Number and energy amplitudes.* The absorption term contains $\int G(\mathbf{x}-\mathbf{X})\cdot\phi_1\,d^3x = \sqrt{2\pi\hbar c}\,[|\nabla|^{1/2}\phi_1](\mathbf{X})$, which is proportional to the energy amplitude $(\hbar c|\nabla|)^{1/2}\phi_1$ of Białynicki-Birula and Sipe [11, 12] at the nucleus. The atom absorbs according to the energy amplitude and not according to the number density at its position, in agreement with photodetection theory [32].

(iv) *Gauge dependence of the photon number.* The transformation (8) displaces the field oscillators by an amount that depends on the electron coordinates. A state without photons in the length gauge contains a cloud of photons in the Coulomb gauge, and the converse holds for the dressed ground state. The photon number of a bound state is therefore not an observable on its own, while the total energy, the electron density and the field measured far from the atom are gauge independent.

(v) *What the photon coordinate means.* The photon has no position operator with localized eigenstates [14], and the photon state vector is often said to have no position-space representation [20]. Eq. (16) does not contradict this, since it is a unitary change of the labels of the one-photon states from modes to points, and $|\phi_1(\mathbf{R};\mathbf{x})|^2$ is a probability density for the label $\mathbf{x}$. It is neither the probability that a point detector at $\mathbf{x}$ responds nor the energy density of the field there; the energy amplitude $(\hbar c|\nabla|)^{1/2}\phi_1$ is local in the field but is not a probability amplitude. Statements about where the photon is refer to the number amplitude and are sharp only on scales larger than the wavelengths that make up the photon; for the scattered packets of Section 3 this scale is $\lambda_L = 134\ nm = 2.5 \times 10^3$ a.u., against a packet width of $1.4 \times 10^4$ a.u.

## 2.5. Truncation at one photon and the polarization energy

We keep the components without and with one photon, $\phi_0(\mathbf{R},t)$ and $\phi_1(\mathbf{R};\mathbf{x},t)$, and set $\phi_2 = 0$. This is Fock's truncation of his hierarchy [10], known from its later use in meson theory as the Tamm–Dancoff approximation [33, 34]:

$$i\hbar\ \partial_t\phi_0 = [H_{el} + V_P]\phi_0 - i\,\mathbf{d}\cdot\int G(\mathbf{x}-\mathbf{X})\cdot\phi_1(\mathbf{R};\mathbf{x})\,d^3x \qquad (19)$$

$$i\hbar\ \partial_t\phi_1 = [H_{el} + \hbar c\,|\nabla_{\mathbf{x}}|]\phi_1 + i\,G(\mathbf{x}-\mathbf{X})\cdot\mathbf{d}\ \phi_0(\mathbf{R}) \qquad (20)$$

Without the RWA the number of excitations is not conserved, and what is truncated is the number of photons present at the same time, not the order in the coupling, and the coupled components contain the emission, reabsorption and re-emission of the photon to all orders. A direct projection of $H'$ onto the two components would keep $V_P$ in both of them. For an atom without a photon, the first-order shift by $V_P$ and the second-order shift by one virtual photon each grow with the cutoff, but by the closure relation over the atomic states their leading growth cancels. In three dimensions the remainder still contains a term quadratic in the cutoff, which by the Thomas–Reiche–Kuhn sum rule is the same for all levels, a linear term, which is removed by mass renormalization, and Bethe's logarithmic term [35]. In the component with one photon the process that would cancel $V_P$ leads to two photons, which have been removed by the truncation, and $V_P$ would remain there as an uncompensated shift of every atom that has emitted, growing with the cutoff. We therefore adopt a sector-dependent prescription and keep $V_P$ only in the component without a photon, where it is compensated within the retained components. Counterterms that depend on the Fock sector are standard in truncated Fock-space methods [36], and Dancoff [34], like Fock, dropped the self-energy from the truncated equations.

The prescription is an approximation motivated by the processes removed with the truncation, and we test it in Section 3 through the cutoff dependence of the entanglement induced by the pulse. A consequence of the prescription is that at large $|y|$ the potential $S_1 y^2$ of the waveguide model (Section 2.6) acts on the component without a photon only.

## 2.6. One-dimensional waveguide model

From here on we use atomic units, $\hbar = m = e = 1$ and $c = 137.036$. The atom is placed in a waveguide with its axis along $x$ that carries a single dispersionless transverse mode, uniform over the cross-section $A_\perp$ at the atom and polarized along $y$. The guide is closed into a ring of length $L$, with the wave numbers $k_m = 2\pi m/L$ and the frequencies $\omega_m = c|k_m|$; the mode $m = 0$ carries no field and is left out. One electron moves along the polarization direction with the coordinate $y$, the nucleus is fixed at $X = 0$, and the dipole is $\mathbf{d} = -y\,\mathbf{e}_y$. The atomic Hamiltonian is the soft-core model

$$h = -\frac{1}{2}\frac{\partial^2}{\partial y^2} - \frac{1}{\sqrt{y^2 + a^2}}, \qquad a^2 = 2 \tag{21}$$

whose ground-state energy is $e_0 = -0.5$, with the first excited state at $e_1 = -0.2329$; the transition frequency is $\omega_{10} = 0.2671$ (7.27 eV, $\lambda_{10} = 171$ nm) and the transition dipole $y_{10} = 1.3$. In the guide the photon amplitude reduces to $\phi_1 = \mathbf{e}_y A_\perp^{-1/2} C_1(y, x, t)$, and we write $P_0(y, t) \equiv \phi_0$. With $\mathbf{d} = -y\,\mathbf{e}_y$, Eqs. (19) and (20) become

$$i\,\partial_t P_0 = [h + F(t)\,y + S_1 y^2] P_0 + i\,y \int G(x)\,C_1(y, x)\,dx \tag{22}$$

$$i\,\partial_t C_1 = [h + F(t)\,y + c\,|\,\partial_x|] C_1 - i\,y\,G(x)\,P_0(y) \tag{23}$$

where $|\,\partial_x|$ multiplies $e^{ikx}$ by $|k|$. The coupling kernel and the constant of the polarization energy are

$$G(x) = \frac{1}{\sqrt{L}} \sum_{m\neq 0} g_m\, e^{ik_m x}, \qquad g_m = \sqrt{\frac{\omega_m\,\Delta k}{A_\perp}}, \qquad \Delta k = \frac{2\pi}{L}, \qquad S_1 = \sum_{m\neq 0} \frac{g_m^2}{\omega_m} = \sum_{m\neq 0} \frac{\Delta k}{A_\perp} \tag{24}$$

Here $g_m$ is the field of one photon of mode $m$ at the atom. $G$ is real and even, with $\int G\,dx = 0$, a positive core at the atom and a negative tail $\propto |x|^{-3/2}$, the one-dimensional counterpart of $|\nabla|^{1/2}\delta^\perp$. The constant $S_1 \simeq 2k_c/A_\perp$ grows linearly with the largest wave number $k_c$ in the sum, while in the guide the remainder of the level shift (Section 2.5) grows only logarithmically, because the density of modes is constant in $k$ and $g_m^2 \propto \omega_m$. No mass counterterm is needed. For the decay $u_1 \to u_0$ of the first excited eigenstate $u_1$ of $h$ to its ground state $u_0$, the golden rule, $\Gamma = 2\pi \sum_m y_{10}^2\, g_m^2\, \delta(\omega_m - \omega_{10})$, gives with Eq. (24) and emission in both directions the rate

$$\Gamma = \frac{4\pi\,\omega_{10}\,y_{10}^2}{c\,A_\perp} \tag{25}$$

and we fix the cross-section $A_\perp$ by choosing $\Gamma$. With $\Gamma = 0.01$, $\Gamma/\omega_{10} \approx 0.04$, the coupling is much stronger than for an atom in free space. It corresponds rather to an engineered emitter in a strongly confining structure, and it makes the electron–photon correlations large enough to be resolved numerically.

The term $F(t)\,y$ couples the electron to the laser pulse, whose classical field $F(t)$ is polarized along $y$. The quantized field starts from its vacuum [22]; the photon number in the guide therefore counts only the scattered quanta, and a scattered photon of any frequency, including the harmonics of the laser, belongs to the one-photon component. Since the gauge transformation (8) commutes with $y$, the pulse enters both components in the same way.

The emission and absorption terms in Eqs. (22) and (23) carry the same real kernel $y\,G(x)$ with opposite imaginary factors, and the total probability $W_0 + W_1$, with the component weights $W_0(t) = \int |P_0|^2\,dy$ and $W_1(t) = \iint |C_1|^2\,dy\,dx$, is conserved. The photon term $c\,|\,\partial_x|$ is not a differential operator and has no local probability current. On the right-moving ($k > 0$) and left-moving ($k < 0$) parts of $C_1$ it reduces to $\mp ic\,\partial_x$, and the photon is transported rigidly at $\pm c$; the two parts overlap only near the atom.

In the waveguide model the guide field is polarized along $y$ and varies only along $x$, while the electron moves along $y$ and the nucleus is fixed, so the dipole approximation is exact. For a real atom the corrections are of relative order $(k\ell)^2$; with $\ell \approx \langle y^2\rangle^{1/2} = 1.34$ this is $7\times10^{-6}$ for photons at $\omega_{10}$ and $2\times10^{-4}$ at the grid cutoff. The uniform transverse profile requires a guide wider than the atom, while Eq. (25) with $\Gamma = 0.01$ gives $A_\perp = 4.1$ a.u.; we therefore regard $A_\perp$ as an effective coupling parameter. Recoil of the atom, a separate source of atom–photon entanglement [17], is excluded by the fixed nucleus.

### 2.7. Electron–photon entanglement: global and local

The state $\{P_0, C_1\}$ is a pure state of the electron and the photon. The state space of the photon consists of the vacuum $|\mathrm{vac}\rangle$ and the one-photon states $|x\rangle$ at the positions $x$, and the state of the system is $|\Psi(t)\rangle = \int P_0(y,t)\,|y\rangle|\mathrm{vac}\rangle\,dy + \iint C_1(y,x,t)\,|y\rangle|x\rangle\,dy\,dx$. Its probability density $|\Psi(t)|^2$ is defined on two sheets of configurations, $|P_0(y,t)|^2$ on the line of configurations without a photon and $|C_1(y,x,t)|^2$ for configurations with one photon, with the total weight $W_0 + W_1 = 1$. The trace over the photon gives the reduced density matrix (RDM) of the electron,

$$\rho_e(y,y') = P_0(y)\,P_0^*(y') + \int C_1(y,x)\,C_1^*(y',x)\,dx \tag{26}$$

and the trace over the electron gives the RDM of the photon, $\rho_\gamma$, with a vacuum block, a one-photon block and coherences between them,

$$\langle\mathrm{vac}|\rho_\gamma|\mathrm{vac}\rangle = W_0, \qquad \langle x|\rho_\gamma|\mathrm{vac}\rangle = \int C_1(y,x)\,P_0^*(y)\,dy,$$
$$\langle x|\rho_\gamma|x'\rangle = \int C_1(y,x)\,C_1^*(y,x')\,dy \tag{27}$$

Since the state is pure, the two RDMs have the same nonzero eigenvalues $\lambda_l$. We use the von Neumann entropy $S_{vN} = -\sum_l \lambda_l \ln\lambda_l$. All entropies refer to the partition between the electron and the transverse field of the length gauge, $\mathbf{D}_\perp$. Since the transformation (8) acts on the electron and the field together, the entropies depend on this choice of partition. Far from the atom $\mathbf{D}_\perp$ coincides with the electric field, and an outgoing photon is described by the same field in both gauges, which does not make the entropies themselves independent of the partition.

When the electron state in the photon component is orthogonal to $P_0$, $\int P_0^*(y)\,C_1(y,x)\,dy = 0$ for all $x$, as it is in the ground state by parity, the coherences in Eq. (27) vanish, and the entropy separates into a photon-number part and a part within the one-photon component,

$$S_{vN} = -W_0\ln W_0 - W_1\ln W_1 + W_1\,S_{vN}\big(C_1/W_1^{1/2}\big) \tag{28}$$

where the last entropy is that of the normalized one-photon component alone.

For the local entanglement we follow the strip construction of Refs. [25, 26]. An ensemble of $M$ walkers samples $|\Psi(t)|^2$, where a walker without a photon ($s_k = 0$) carries an electron coordinate $y_k$ and is distributed with $|P_0(y_k)|^2$, and a walker with a photon ($s_k = 1$) carries the pair $(y_k, x_k)$ and is distributed with $|C_1(y_k,x_k)|^2$. The electron wave conditioned on walker $k$ is the slice of the state at the photon configuration of the walker,

$$\varphi_k(y) = \begin{cases} P_0(y)/W_0^{1/2}, & s_k = 0 \\ C_1(y, x_k)/n(x_k)^{1/2}, & s_k = 1 \end{cases} \tag{29}$$

where $n(x) = \int |C_1(y, x)|^2\, dy$ is the photon number density, and the photon state conditioned on walker $k$ is the slice at its electron coordinate,

$$|\chi_k\rangle = \frac{P_0(y_k)\,|\mathrm{vac}\rangle + \int C_1(y_k, x)\,|x\rangle\, dx}{\rho_e(y_k, y_k)^{1/2}} \tag{30}$$

which contains the vacuum amplitude next to the one-photon wave. A strip $m$ is a range of the coordinate of the particle whose RDM we want. An electron strip contains the $M_m$ walkers with $y_k$ in it, $M_m^{(0)}$ of them without a photon, and its RDM is the average of the projectors on the electron waves (29),

$$\rho_m^{(e)}(y, y') = \frac{1}{M_m}\sum_{k\in m} \varphi_k\,(y)\,\varphi_k^*(y') = \frac{1}{M_m}\left[M_m^{(0)}\,\frac{P_0(y)\,P_0^*(y')}{W_0} + \sum_{k\in m_1} \frac{C_1(y, x_k)\,C_1^*(y', x_k)}{n(x_k)}\right] \tag{31}$$

where $m_1$ denotes the walkers of the strip with a photon. A photon strip contains the $M_m$ walkers with $x_k$ in it, the walkers without a photon forming one additional domain, and its RDM is the average of the projectors on the photon states (30),

$$\rho_m^{(\gamma)} = \frac{1}{M_m}\sum_{k\in m} |\,\chi_k\rangle\langle\chi_k| \tag{32}$$

Its elements $\langle \mathrm{vac}|\rho_m^{(\gamma)}|\mathrm{vac}\rangle$, $\langle x|\rho_m^{(\gamma)}|\mathrm{vac}\rangle$ and $\langle x|\rho_m^{(\gamma)}|x'\rangle$ are the averages over the walkers of the strip of $|P_0(y_k)|^2$, $C_1(y_k, x)\,P_0^*(y_k)$ and $C_1(y_k, x)\,C_1^*(y_k, x')$, each divided by $\rho_e(y_k, y_k)$, and have the structure of the global photon RDM (27). The local von Neumann entropy of a strip is

$$S_{vN}^{(m)} = -\mathrm{Tr}\,\rho_m \ln \rho_m \tag{33}$$

with $\rho_m = \rho_m^{(e)}$ or $\rho_m^{(\gamma)}$. Each strip RDM is an average of normalized projectors and is therefore positive, with unit trace. It describes the particle conditioned on its position in the strip, as after a weak measurement of the position followed by postselection [26], and its entropy measures the entanglement between the particle localized in the strip and the other particle. For a product state all conditional waves coincide and every strip entropy vanishes; the interaction makes the conditional waves of the walkers in a strip differ from one another, and the strip entropy grows with this diversity. This local entanglement of one particle with the whole of the other is not the entanglement between two regions of space. When all walkers belong to one strip, Eqs. (31) and (32) reproduce the global RDMs (26) and (27) for a large number of walkers. If the one-photon component is a product, $C_1 = b(y)f(x)$ with $b \perp P_0$, all walkers with a photon carry the same electron wave, the RDM of an electron strip has two orthogonal blocks with the weights $f_m = M_m^{(0)}/M_m$ and $1 - f_m$, and its von Neumann entropy is the binary entropy $S_{vN}^{(m)} = -f_m \ln f_m - (1 - f_m)\ln(1 - f_m)$; the local entropy records how the photon number is mixed locally. When the electron state that goes with the photon depends on the photon position, the electron waves in Eq. (31) differ from walker to walker, and the strip entropy acquires a spatial structure of its own.

## 3. Numerical implementation and results

Both coordinates are discretized on uniform periodic grids, the electron with $N_y = 128$ points and $\Delta y = 0.4$, the photon with $N_x = 1024$ points in the box $L = 2.4\,c \times 900$, which allows propagation up to $t = 900$ a.u., and all derivatives are taken spectrally by fast Fourier transform. The photon grid spacing $\Delta x = 289$ a.u. sets the largest coupled wave number, $k_c = \pi/\Delta x = 1.087 \times 10^{-2}$ ($c\,k_c = 1.49$). The atomic eigenstates are obtained by diagonalizing $h$ on the electron grid; they provide $u_0$ and the values $\omega_{10}$ and $y_{10}$, from which Eq. (25) with $\Gamma = 0.01$ gives the cross-section $A_\perp = 4.137$. The coupling constants $g_m$ of Eq. (24) are set on the photon grid, with the Nyquist mode uncoupled to keep $G$ real, and $G(x)$ follows by one inverse FFT; its norm is $\| G \| = (\int G^2\,dx)^{1/2} = 0.0625$.

The time evolution uses symmetric (Strang) splitting of the propagator over the time step $\Delta t = 0.05$. The kinetic energy of the electron and the photon energy $c|\,\partial_x|$ are applied as phase factors in Fourier space, half a step before and half a step after the other terms, and the potential $V(y,t) = -(y^2 + a^2)^{-1/2} + F(t)\,y$, the same in both components, as a phase factor on the grid. The polarization energy $S_1 y^2$ acts only on $P_0$ and is applied in two half steps around the coupling step. The coupling terms are the only ones that are neither local on the grid nor diagonal in Fourier space; at each grid point $y_i$ they connect the amplitude $P_0(y_i)$ with the whole photon wave $C_1(y_i,\cdot)$. Keeping only these terms in Eqs. (22) and (23), and writing the amplitude of the photon wave along the normalized kernel $\hat{G} = G/\| G \|$ as $\xi(y_i) = \int \hat{G}(x)\,C_1(y_i,x)\,dx$, we obtain

$$\partial_t P_0(y_i) = y_i\ \| G \|\ \xi(y_i), \qquad \partial_t \xi(y_i) = -y_i\ \| G \|\ P_0(y_i) \tag{34}$$

while the part of $C_1(y_i,\cdot)$ orthogonal to $\hat{G}$ does not change. At each $y_i$ the coupling therefore acts only in the two-dimensional space of the state without a photon and the one-photon state of shape $\hat{G}$, where Eq. (34) describes a rotation of the two amplitudes with the angular velocity $y_i \| G \|$. Its exact solution over the step is

$$P_0 \leftarrow \cos\theta_i\,P_0 + \sin\theta_i\,\xi, \qquad C_1(y_i,x) \leftarrow C_1(y_i,x) + [(\cos\theta_i - 1)\,\xi - \sin\theta_i\,P_0]\hat{G}(x) \tag{35}$$

where $\theta_i = y_i \| G \| \Delta t$, and $P_0$ and $\xi$ on the right are the values before the step. The step is exact and unitary, conserves $W_0 + W_1$ to round-off, and costs one projection and one update along $\hat{G}$ per grid point. An absorbing mask over the outer 6 a.u. of the electron grid acts on both components after each step.

The dressed ground state is obtained with the same steps in imaginary time, $t \to -i\tau$, where the phase factors become decay factors, the rotation (35) becomes the corresponding hyperbolic transformation, and the total norm is restored after each step. Starting from $P_0 = u_0$, $C_1 = 0$, the relaxation converges to the lowest state of Eqs. (22) and (23) without the pulse, the atom together with the photon component bound to it. The relaxed state is then propagated with the pulse $F(t)$ from $t = 0$. The pulse is Gaussian, centred at $t_0 = 156.9$, with the peak field $F_0 = 0.0534$ (peak intensity $1.0 \times 10^{14}$ W/cm$^2$), an intensity FWHM of 2.24 fs and the carrier frequency $\omega_L = 0.340$ ($\lambda_L = 134$ nm, 9.25 eV), off-resonant between the lines $u_0 \to u_1$ (7.27 eV) and $u_0 \to u_3$ (11.3 eV) and below the ionization energy of 13.6 eV.

For the profiles in Figure 3, $M = 2 \times 10^4$ walkers sample $|\Psi(t)|^2$ by the Metropolis algorithm. For the strip entropies we sample $M = 2 \times 10^7$ walkers independently from $|\Psi(t)|^2$ on the grids of both components, divide them into strips and form the RDMs (31) and (32) with the strict conditional waves (29) and (30), slices of the exact solution at the walker positions. On the grid the photon state

(30) is the vector $\left(P_0(y_k),\, \Delta x^{1/2} C_1(y_k, x_1), \ldots, \Delta x^{1/2} C_1(y_k, x_{N_x})\right)/\rho_e(y_k, y_k)^{1/2}$ in the orthonormal basis of the vacuum and the grid states $\Delta x^{1/2}|x_l\rangle$. Walkers in the same grid cell carry the same conditional wave, so the sums over the walkers are taken cell by cell with the number of walkers in each cell. The statistical error of the strip entropies decreases as $M^{-1/2}$; at $M = 2 \times 10^7$ it is about $5 \times 10^{-4}$ in the central electron strips.

The propagation was checked in two ways. Halving the time step to $\Delta t = 0.025$ changes the global entropy by less than $2 \times 10^{-4}$ at all times. Starting from the bare excited state $u_1$ without a pulse, the atom decays with the rate 0.0099, against the golden-rule value $\Gamma = 0.01$ of Eq. (25).

We now apply the method to the atom of Section 2.6 driven by the laser pulse. The propagation stops at the end of the pulse, $t = 7.6\, fs$, and the photon cutoff is $c\, k_c = 1.49$ unless stated otherwise. We report von Neumann entropies throughout. The relaxation in imaginary time gives the dressed ground state with the one-photon weight $W_1 = 0.0062$. The electron state of the one-photon component is odd in $y$, while $P_0$ is even, and the decomposition (28) applies. The photon-number part $-W_0 \ln W_0 - W_1 \ln W_1 = 0.0374$ accounts for nearly the whole global entropy $S_{vN} = 0.0375$, and the entanglement within the one-photon component adds only $2 \times 10^{-5}$. Figure 1(a) shows the joint probability density $|C_1(y,x)|^2$ of the electron at $y$ and the (virtual) photon at $x$. This component is not an emitted photon but the photon cloud of the dressed field vacuum, the part of the dressed ground state in which the electron and the field are correlated, and it carries no energy away. How large it appears depends on the field variable used to count photons (Section 2.4), while the dressed state itself, its energy and the field far from the atom do not. It forms a narrow column at $x = 0$, split into two lobes in $y$ with a node at $y = 0$, where the coupling $y\, G(x)$ of Eqs. (22) and (23) vanishes. The truncation at one photon suffices for this dressing, because the counter-rotating part of the coupling connects the atom without a photon directly with the one-photon component.

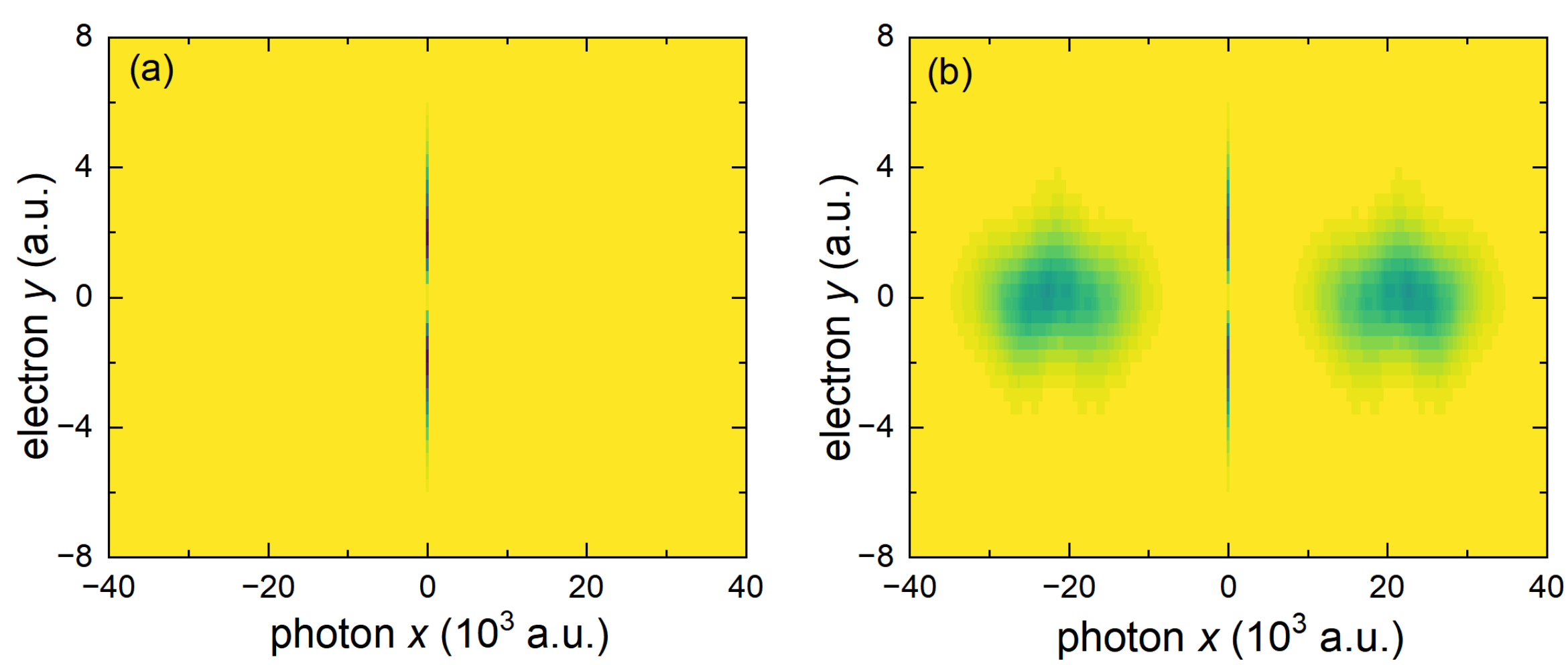

**Figure 1.** Joint probability density $|C_1(y,x)|^2$ of the electron at $y$ and the photon at $x$ in the one-photon component, on a common linear colour scale from yellow (zero) to dark blue (maximum), (a) at $t = 0$, before the pulse (ground state dressed by the field vacuum), and (b) at the end of the pulse. The weight of the one-photon component is $W_1 = 0.0062$ in (a) and 0.175 in (b). The column at $x = 0$ is the vacuum field correlated with the electron; the two lobes in (b) are the two packets of the scattered photon moving along the guide.

Figure 1(b) shows the joint distribution at the end of the pulse. The scattered photon has separated into two packets that move along the guide with the speed of light in opposite directions. They are centred at $|x| = 2.17 \times 10^4$, with a full width at half maximum of $1.4 \times 10^4$, of the order of $c$ times the duration of the pulse, while the column at the atom has returned close to its ground-state form and carries 3% of $W_1$. Within each packet the electron is distributed as in the component without a photon, and its distribution does not depend on the position of the photon.

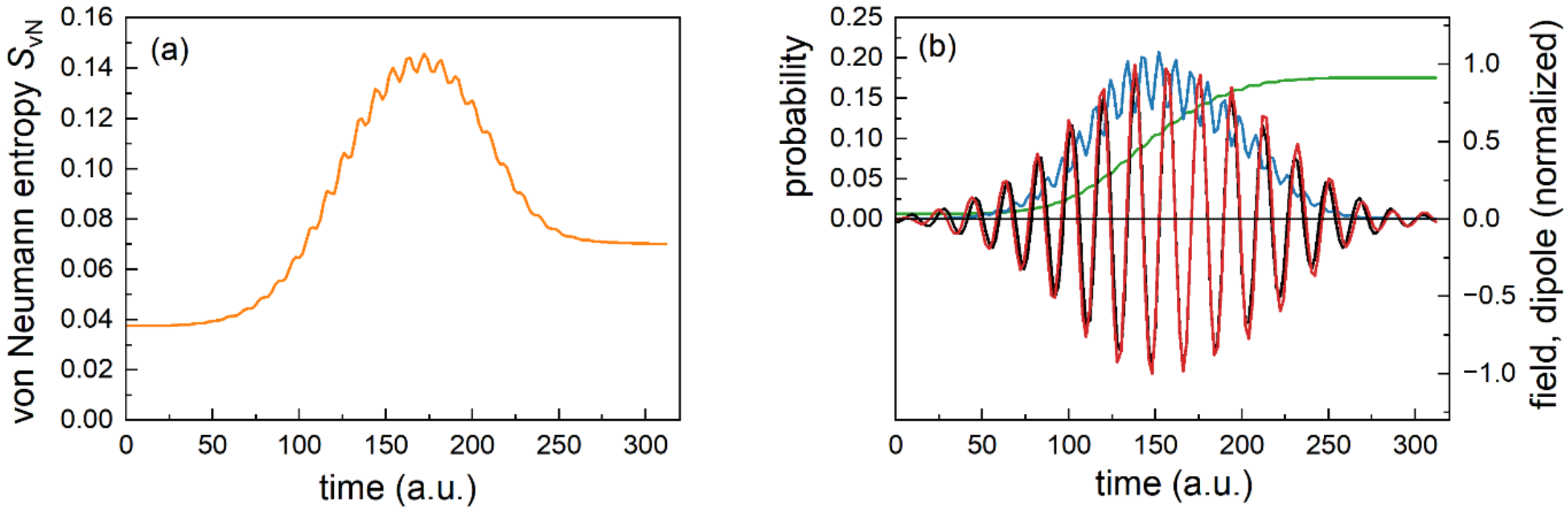

**Figure 2.** (a) Global von Neumann entropy of the electron–photon state. (b) One-photon weight $W_1$ (green) and probability of the excited atom without a photon, $W_0 - |\langle u_0|P_0\rangle|^2$ (blue), left axis; laser field $F(t)$ (black) and dipole moment $\langle y\rangle(t)$ of the dressed atom (red), each divided by its maximum absolute value, right axis.

The global von Neumann entropy (Figure 2a) rises from $0.0375$ to $0.146$ at $t = 172$, shortly after the maximum of the pulse, and falls to $0.070$ at the end, although the photon weight stays at $0.175$, so the entanglement between the electron and the photon is mostly transient. The entropy and the probability of the excited atom (Figure 2b) fall together after the maximum of the pulse, when the part of the electron cloud that the field has driven away from the nucleus returns to the core (recombination). Since the absorber removes less than $10^{-3}$ of the probability, the returning part is the polarized bound cloud rather than ionized electrons, and the atom ends close to its dressed ground state whether a photon was scattered or not. Equation (28) shows why this removes most of the entanglement. If the electron states of the two components were orthogonal, as in the ground state, the photon-number part alone would give $-W_0 \ln W_0 - W_1 \ln W_1 = 0.46$ at the end. After the return to the core, however, the electron state of the one-photon component nearly coincides with that of the component without a photon. With the prescription of Section 2.5 the change of the entropy induced by the pulse converges for cutoffs two and three times larger, obtained with shorter boxes of the same number of points, while keeping the polarization energy also in the component with one photon makes it grow with the cutoff.

Figure 2(b) shows the laser field and the dipole moment $\langle y\rangle$ of the dressed atom, each normalized to its maximum, together with the populations. The dipole follows the field cycle by cycle, with a small phase lag, and returns to zero with the pulse. This is the response of an atom driven off-resonance, which is polarized by the field but not left in an excited state. Accordingly, the probability that the atom is excited while no photon is present, $W_0 - |\langle u_0|P_0\rangle|^2$, follows the envelope with a ripple at twice the laser frequency, since it is quadratic in the field, reaches $0.21$ at the maximum of the pulse and falls to $3 \times 10^{-4}$ at the end. The one-photon weight $W_1$ grows during the pulse and saturates at $0.175$, as the driven dipole scatters a photon into the guide with growing probability. Since 3% of $W_1$ remains in the photon cloud at the atom (Figure 1b), the mean number of photons emitted by the atom during the pulse is $\bar{n} \approx 0.17$. If the number of scattered photons obeyed Poisson statistics, the probability of two scattered photons, which the truncation omits, would be about $\bar{n}^2/2 \approx 0.014$; this is an estimate under that assumption, not a bound.

Figure 3 compares the spatial profiles of the electron and of the photon at the end of the pulse, obtained from the solution of Eqs. (22) and (23), with kernel density estimates (KDE) over the Metropolis walkers. The electron density with and without a photon, the electron density in the one-photon component and the photon profile of the packets are reproduced within the statistical noise. The two electron profiles nearly coincide (Figure 3a). The core of the photon distribution at the atom is narrower than the kernel and is not resolved by the estimate.

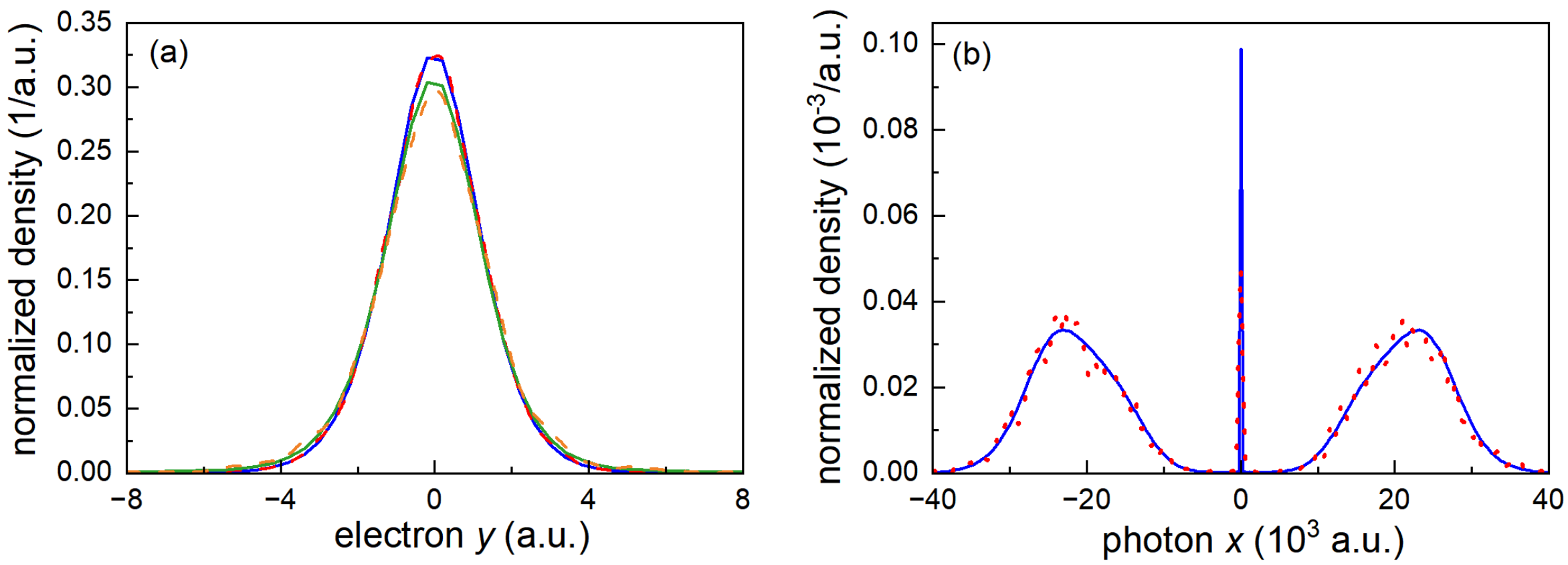

**Figure 3.** Normalized spatial profiles at the end of the pulse, from the solution of Eqs. (22) and (23) (solid lines) and from kernel density estimates over the walkers (broken lines). (a) Electron density with and without a photon (blue solid; estimate over all walkers, red dashed) and in the one-photon component, $\int |C_1|^2 dx$ (green solid; estimate over the walkers with a photon, orange dashed). (b) Photon number density $n(x)$ (blue solid; estimate over the walkers with a photon, red dotted); the narrow peak at the atom is the photon cloud of the dressed vacuum. All profiles are normalized to one.

Figure 4 shows the local von Neumann entropies of the strips at the end of the pulse and in the dressed ground state. The two particles show opposite patterns. The electron entropy has its minimum at the nucleus (0.043) and grows with the distance from it, to 0.48 at $|y| = 6$, while the photon entropy is largest near the atom, $|x| \lesssim 10^4$, where photons are created and absorbed (0.27), and smaller in the outgoing packets (0.09 to 0.14). The cloud at the atom is one photon grid cell wide (Figure 3b), so the photon entropy there is an average over this cell, whereas the outgoing packets are resolved. Both patterns originate in the dressed ground state, in which the photon is confined to the cloud at the atom (strip entropy 0.12, not shown in Figure 4b), and they follow from the form of the coupling. Equations (22) and (23) contain no potential that depends on the distance between the electron and the photon; their coordinates do not even refer to the same direction. The coupling converts the component without a photon into the one-photon component and back, with an amplitude that factorizes into the dipole lever arm $y$ of the electron and the field profile $G(x)$ at the atom.

To first order in the coupling, the photon state conditioned on the electron at $y_k$, Eq. (30), consists of the vacuum and a photon cloud at the atom whose amplitude is proportional to the electron position. Its one-photon part changes sign with $y_k$; electrons on the two sides of the nucleus come with photon clouds of opposite sign, and the RDM of a photon strip at the atom mixes these different states. A photon in an outgoing packet has left the region in which $G$ acts. Its shape was fixed at the time of emission, and after coherent scattering the electron state that goes with it does not depend on the position of the photon, as seen in Figure 1(b) and Figure 3(a). The local entanglement of the photon is therefore concentrated where photons are created and absorbed, and that of the electron where its dipole coupling is strong.

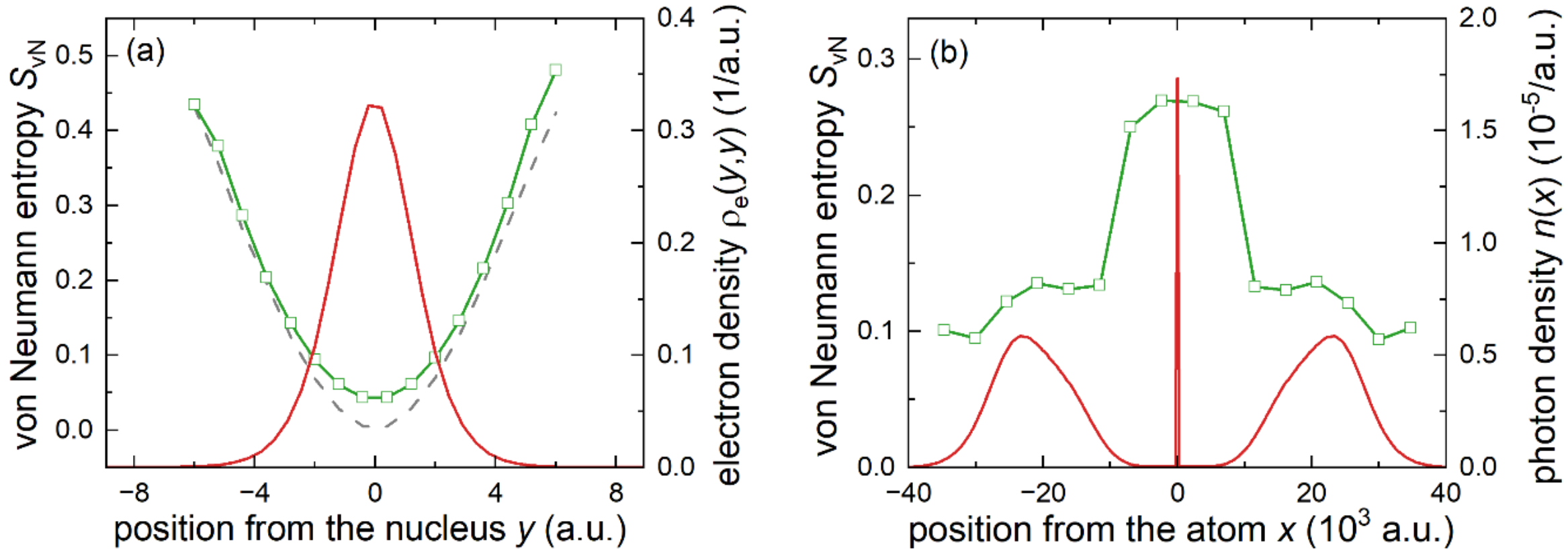

**Figure 4.** Local von Neumann entropies of the strips. (a) Electron strips at the end of the pulse (green) and in the dressed ground state at $t = 0$ (grey, dashed), as a function of the position $y$ measured from the nucleus; electron density (red, right axis). (b) Photon strips at the end of the pulse (green), as a function of the position $x$ along the guide, measured from the atom; photon number density $n(x)$ (red, right axis).

## 4. Conclusions

We have written the photon-number hierarchy of nonrelativistic QED in the configuration space of the photons, in the length gauge, starting from Fermi's oscillator representation of the field. In this form the photon has a coordinate of its own and is created around the atom with the nonlocal kernel $|\nabla|^{1/2}\delta^{\perp}$, and the state of the electrons and the field is a sequence of wave functions in configuration spaces whose dimension changes with the photon number. Truncated at one photon, the equations call for a sector-dependent treatment of the polarization energy, which we keep only in the component without a photon; with this approximation the entanglement induced by the pulse converges with the cutoff. We solved the equations without the rotating-wave approximation for a one-dimensional model atom in a waveguide with a single transverse mode, driven by a short off-resonant laser pulse, with the electron, including its continuum, and the photon on real-space grids.

During the pulse the atom emits a photon into the guide with probability 0.17. The global electron–photon entanglement rises from its value in the dressed ground state to a maximum shortly after the peak of the pulse and then falls, because the polarized electron cloud returns to the core and the electron states that accompany the components with and without a photon become nearly the same. After the pulse the state is close to a product of the electron state and a superposition of the field vacuum with one scattered photon. The local entanglement, obtained from walkers sampled over both photon-number components, shows opposite patterns for the two particles. The electron entropy is smallest at the nucleus and grows with the distance from it, as a map of the local strength of the dipole coupling, while the photon entropy is largest near the atom, where photons are created and absorbed, and small in the outgoing packets, whose shape was fixed at the time of emission.

With the basic equations and the solution method established and tested here, the primary next goal is to extend the number of atoms and photons to realistic values within time-dependent quantum Monte Carlo, in which the photons, like the electrons, are represented by walkers and guide waves and the photon number becomes an additional variable of the walker; chains and lattices of atoms coupled to a common field are natural first systems. Further extensions are two-dimensional geometries, in which the photon cloud acquires the full tail of $|\nabla|^{1/2}\delta^{\perp}$, and the two-photon component, which becomes necessary when the mean number of scattered photons approaches one.

**Acknowledgments**

This research is based upon work supported by the Bulgarian Ministry of Education and Science as a part of National Roadmap for Research Infrastructure, grant D01-351/13.12.2023 (ELI ER-IC BG), and by the National Science Fund, grant КП-06-Н78/6.